# Passive Taxonomy of Wifi Clients using MLME Frame Contents

Denton Gentry
dgentry@google.com
denny@geekhold.com

Avery Pennarun
apenwarr@google.com
apenwarr@gmail.com

*In supporting Wifi networks it is useful to identify the type of client device connecting to an AP. Knowing the type of client can guide troubleshooting steps, allow searches for known issues, or allow specific workarounds to be implemented in the AP. For support purposes a passive method which analyzes normal traffic is preferable to active methods, which often send obscure combinations of packet options which might trigger client bugs.*

*We have developed a method of passive client identification which observes the contents of Wifi management frames including Probes and Association requests. We show that the management frames populated by modern Wifi chipsets and device drivers are quite distinguishable, making it possible in many cases to identify the model of the device. Supplementing information from the Wifi management frames with additional information from DHCP further extends the set of clients which can be distinguished.*

## Terminology

We define Taxonomy as identifying the type of a client device without identifying an individual user. For example, taxonomy would distinguish a smartphone from Vendor A versus Vendor B, but not identify a unique individual device nor provide information to follow a device as it travels from one place to another.

In earlier literature this technique has generally been referred to as Fingerprinting, but in recent years the term Fingerprinting has evolved to mean the identification of an individual. We do not see value in attempting to reverse this evolution in the usage of the term, as it has occurred organically with improvements in the fine distinctions which these mechanisms can detect. Instead we believe it will be more productive to use a distinct term.

In Biology, taxonomy concerns identification of species. We propose to consider the model of a device as a rough equivalent of a species, and use the term Taxonomy for the identification mechanism.

## Overview of Proposed Taxonomy Approach

With the mainstream adoption of 802.11n and 802.11ac, Wifi management frames contain a rich set of optional fields, capability bitmasks, and other information which vary substantially between different Wifi devices. By listing (in order) the parameters present in several common types of management frames, and extracting a few specific bitmasks from these frames, a highly specific signature can be developed.

This signature is most strongly influenced by the chipset, which determines the values populated in the various capability bitmasks. It is next most strongly influenced by the device driver and wifi software stack, which determine the specific Information Elements present. Finally, a few values relating to power levels and number of antennas are determined by the PCB board design. The combination of all of these can generally identify the model of the device.

## Introduction to MLME

The Wifi MAC Layer Management Entity (MLME) comprises a number of different types of packets used in the operation of the Wifi network.

Most Wifi MLME frames consist of a set of Fixed Parameters which are always present followed by Tagged parameters, optional fields which are implemented as Type-Length-Value tuples. The standards documents define many parameters, and there is a vendor extension mechanism for private entities to add their own parameters.

For example:

```
IEEE 802.11 wireless LAN management frame
    Tagged parameters (78 bytes)
        Tag: SSID parameter set: Broadcast
            Tag Number: SSID parameter set (0)
            Tag length: 0
        Tag: Supported Rates
            Tag Number: Supported Rates (1)
            Tag length: 4
        Tag: Extended Supported Rates
            Tag Number: Extended Supported Rates (50)
            Tag length: 8
        Tag: HT Capabilities (802.11n D1.10)
            Tag Number: HT Capabilities (802.11n D1.10) (45)
            Tag length: 26
        Tag: Vendor Specific: Epigram: HT Capabilities (802.11n D1.10)
            Tag Number: Vendor Specific (221)
            Tag length: 30
            OUI: 00-90-4c (Epigram)
            Vendor Specific OUI Type: 51
```

There are a large number of frame types described as part of MLME. For the purposes of this paper, we will focus on two.

**Probe** frames are sent by clients searching for an access point, either a specific AP or broadcasting to find any available APs. The client probe will generally include rather a lot of information about its own capabilities including supported rates and encodings, authentication capabilities, and its support for higher speed operation as defined in 802.11n and 802.11ac.

**Association** frames are sent by clients to ask the AP to add the client to the wireless LAN. As with Probe frames, Association frames generally include substantial information about the capabilities of the client requesting the association.

The number of optional parameters has expanded over time via new Wifi specifications, especially starting with the widely implemented IEEE802.11n-D2.0 in 2007. Wifi clients developed since then have gradually added more parameters in the Probe and Association frames they send.

## Proposed Taxonomy Approach

The management entity in an Access Point handles processing of MLME frames received from other stations. As part of its handling of Probe Request and Association Request frames, the AP examines the series of Tagged Parameters seen from each client, extracting information which is used to make a signature. This information is concatenated into a simple text string.

For example, these are the signatures for iPhone 6s and Nexus 6P in 5GHz operation:

**iPhone 6s**
```
wifi4|probe:0,1,45,127,107,191,221(0050f2,8),221(001018,2),htcap:00
6f,htagg:17,htmcs:0000ffff,vhtcap:0f815832,vhtrxmcs:0000fffa,vhttxm
cs:0000fffa,extcap:0400088400000040|assoc:0,1,33,36,48,70,45,127,19
1,221(001018,2),221(0050f2,2),htcap:006f,htagg:17,htmcs:0000ffff,vh
tcap:0f815832,vhtrxmcs:0000fffa,vhttxmcs:0000fffa,txpow:e002,extcap
:0400000000000040
```

**Nexus 6P**
```
wifi4|probe:0,1,45,191,221(0050f2,4),221(506f9a,9),221(001018,2),ht
cap:006f,htagg:17,htmcs:0000ffff,vhtcap:0f815832,vhtrxmcs:0000fffa,
vhttxmcs:0000fffa,wps:Nexus_6P|assoc:0,1,33,36,48,45,191,221(001018
,2),221(0050f2,2),htcap:006f,htagg:17,htmcs:0000ffff,vhtcap:0f81583
2,vhtrxmcs:0000fffa,vhttxmcs:0000fffa,txpow:e002
```

In the signature string:
- "probe:0,1,45,127" etc are the numeric identifiers of the Tagged Information Elements in a Probe Request, in the order in which they appear.
- "assoc:0,1,33,36," etc are the identifiers of the Tagged Information Elements in an Association Request, in the order in which they appear.
- "221(0050f2,8)" is a Vendor-Specific IE for vendor 00:50:f2 (Microsoft), with subtype 8. "221(001018,2)" is a Vendor-Specific IE for vendor 00:10:18 (Broadcom), with subtype 2. Etc, etc.
- "htcap" is the capabilities bitmask from the optional HT Capabilities Information Element, if it is present. The HT Capabilities IE was added in 802.11n.
- "htagg" is the A-MPDU Parameters bitmask from an HT Capabilities Information Element.
- "htmcs" is the RX Supported Modulation and Coding Scheme bitmask from an HT Capabilities Information Element.
- "vhtcap" is the capabilities bitmask from the optional VHT Capabilities Information Element, if it is present. The VHT capabilities IE was added in 802.11ac.
- "vhtrxmcs" and "vhttxmcs" are the RX MCS Map and TX MCS Map from the VHT Supported MCS Set field from a VHT Capabilities Information Element.
- "extcap" is the Extended Capabilities IE, which numerous 802.11 standards have added to. It varies in length depending on what the client implements.

- "txpow" is the minimum and maximum power values from a Power Capability IE. The Power Capability IE was added in 802.11h. This is tied to a particular board design and its choice of antennas and power amplifiers, and is very helpful in distinguishing devices which use the same chipset and OS. At the time of this writing in early 2016 this is especially useful for Apple and Samsung devices, which are quite common and which tend to use the same software on a variety of different models using a small variety of Wifi chipsets.
- "wps" is the model name from a Wifi Protected Setup IE, if present. There are a few common clients which include WPS headers at all times, and the model name is highly distinctive to that species of client. The signature implementation only allows alphanumeric characters, other characters are replaced by an underscore.

## Distinctiveness of Signatures

The signature reflects a combination of the specific wifi chipset, device driver, WPA supplicant, and PCB layout of the client device. Devices which use the same Wifi chipset but different software will have similarities in their signatures, yet often have sufficient differences allowing them to be distinguished.

For example, the LG G4 running Android 5.1 and the iPhone 6 running iOS 9.2 both use the BCM4339 chipset. The signatures are shown below, with bold text and whitespace added to illustrate where they differ.

**LG G4**
```
wifi4|probe:0,1,3,45,127,107,191,221(506f9a,16),         221(0
01018,2),221(00904c,51),221(00904c,4),221(0050f2,8),htcap:016f,htag
g:17,htmcs:000000ff,vhtcap:0f805932,vhtrxmcs:0000fffe,vhttxmcs:0000
fffe,extcap:0000088001400040|assoc:0,1,33,36,48,  45,127,191,221(0
01018,2),221(00904c,4),221(0050f2,2),htcap:016f,htagg:17,htmcs:0000
00ff,vhtcap:0f805932,vhtrxmcs:0000fffe,vhttxmcs:0000fffe,txpow:1d01
,extcap:000000800140040
```

**iPhone 6**
```
wifi4|probe:0,1,   45,127,107,191,              221(0050f2,8),221(0
01018,2),                                        htcap:0063,htag
g:17,htmcs:000000ff,vhtcap:0f805032,vhtrxmcs:0000fffe,vhttxmcs:0000
fffe,extcap:040008840000040|assoc:0,1,33,36,48,70,45,127,191,221(0
01018,2),        221(0050f2,2),htcap:0063,htagg:17,htmcs:0000
00ff,vhtcap:0f805032,vhtrxmcs:0000fffe,vhttxmcs:0000fffe,txpow:e002
,extcap:040000000000040
```

Even devices which are much more similar can nonetheless have distinct signatures. For example consider the iPhone 5 and 5s, both 802.11n devices using the Broadcom 4334 chipset and running iOS. The signatures are very similar, but iPhone 5s advertises support for BSS Transition in its Extended Capabilities IE and the values in the Power Capability IE differ between the two devices owing to their differing board designs. iPhone 5 may add BSS

Transition support in a future iOS update, but the Power Capabilities should remain distinct forever as it is tied to the board design not software.

**iPhone 5**
```
wifi4|probe:0,1,45,127,107,221(001018,2),221(00904c,51),221(0050f2,
8),htcap:0062,htagg:1a,htmcs:000000ff,extcap:000000004|assoc:0,1,33,
36,48,45,221(001018,2),221(00904c,51),221(0050f2,2),htcap:0062,htag
g:1a,htmcs:000000ff,txpow:1504
```

**iPhone 5s**
```
wifi4|probe:0,1,45,127,107,221(001018,2),221(00904c,51),221(0050f2,
8),htcap:0062,htagg:1a,htmcs:000000ff,extcap:00000804|assoc:0,1,33,
36,48,45,221(001018,2),221(00904c,51),221(0050f2,2),htcap:0062,htag
g:1a,htmcs:000000ff,txpow:1603
```

## Supplemental Information: OUI

Signatures of devices have become more distinctive over time as successive Wifi standards have added ever more optional elements, and as vendors seek to differentiate their offerings with vendor-specific extensions.

In a year of working with this signature mechanism we have found devices where the information extracted from Wifi MLME frames is insufficient to distinguish between them. For example consider the Moto E, Sony Xperia Z Ultra, and Oneplus X:

**Moto E (2nd gen)**
```
wifi4|probe:0,1,50,3,45,221(0050f2,8),htcap:012c,htagg:03,htmcs:000
000ff|assoc:0,1,50,33,48,70,45,221(0050f2,2),127,htcap:012c,htagg:0
3,htmcs:000000ff,txpow:170d,extcap:00000a0200000000
```

**Sony Xperia Z Ultra**
```
wifi4|probe:0,1,50,3,45,221(0050f2,8),htcap:012c,htagg:03,htmcs:000
000ff|assoc:0,1,50,33,48,70,45,221(0050f2,2),127,htcap:012c,htagg:0
3,htmcs:000000ff,txpow:170d,extcap:00000a0200000000
```

**Oneplus X**
```
wifi4|probe:0,1,50,3,45,221(0050f2,8),htcap:012c,htagg:03,htmcs:000
000ff|assoc:0,1,50,33,48,70,45,221(0050f2,2),127,htcap:012c,htagg:0
3,htmcs:000000ff,txpow:170d,extcap:00000a0200000000
```

The Wifi portion of these signatures is identical. They use the same chipset with the same Android OS and driver in a very similar board design. These devices are added to the database with a qualifier, that the OUI of their MAC address be that of Motorola or Sony or Oneplus, respectively.

## Supplemental Information: DHCP

Some time ago engineers at inverse.ca developed a signature mechanism to identify the operating system of DHCP clients, by listing the Options present in the DHCP Request in the order in which they appear. They publish a database of known DHCP signatures called fingerbank (https://fingerbank.inverse.ca/). This DHCP signature mechanism inspired the Wifi MLME mechanism described in this paper.

Some examples of DHCP signatures:
- **Android:** `1,33,3,6,15,26,28,51,58,59`
- **Chrome OS:** `1,121,33,3,6,12,15,26,28,51,54,58,59,119`
- **iOS:** `1,3,6,15,119,252`

We use DHCP identification to supplement the Wifi information in cases where the signature is indistinct. The Roku HD 2500, Withings Scale, and Amazon Dash Button are three examples where the Wifi signature is the same but the DHCP signature allows them to be distinguished.

**Roku HD 2500:**
```
wifi4|probe:0,1,50,45,3,221(001018,2),221(00904c,51),htcap:110c,htagg:19,ht
mcs:000000ff|assoc:0,1,48,50,45,221(001018,2),221(00904c,51),221(0050f2,2),
htcap:110c,htagg:19,htmcs:000000ff
dhcp|1,3,6,15,12
```

**Withings Scale**
```
wifi4|probe:0,1,50,45,3,221(001018,2),221(00904c,51),htcap:110c,htagg:19,ht
mcs:000000ff|assoc:0,1,48,50,45,221(001018,2),221(00904c,51),221(0050f2,2),
htcap:110c,htagg:19,htmcs:000000ff
dhcp|1,3,28,6
```

**Amazon Dash Button**
```
wifi4|probe:0,1,50,45,3,221(001018,2),221(00904c,51),htcap:110c,htagg:19,ht
mcs:000000ff|assoc:0,1,48,50,45,221(001018,2),221(00904c,51),221(0050f2,2),
htcap:110c,htagg:19,htmcs:000000ff
dhcp|1,3,6
```

## Supplemental Heuristics: Hostname and DNS-SD

Even with the supplemental information to disambiguate similar devices, we have a few cases where signatures from different models built from the same components by the same manufacturer display the same signature in their MLME frames. For example, the 2nd generation iPad Air, 4th generation AppleTV, and iPhone 6s/6s+ have the same signature in their Wifi MLME frames.

In these cases the taxonomic identification is uncertain. We rely on additional information outside of the scope of this mechanism, such as the hostnames included in the DHCP Request or in information advertised via DNS-SD, to determine the model of the device.

## Qualifying Signatures

In the time we have been working with this mechanism we have also found that if entries are added to the database which alias several distinct devices, it is exceedingly difficult to figure out after the fact and correct it. We have therefore become more cautious, and apply

distinctiveness criteria to signatures in the database:

- if the signature contains a descriptive WPS model name, we consider that to be sufficient by itself to ensure the signature is distinct.

    We have to add a caveat of *descriptive* model name because early in WPS deployment it was common practice for clients to include a model name of a single space, in order to work around bugs in some early AP implementations which did not handle an empty model name[1]. Devices with non-descriptive WPS model names in their MLME frames are still out there, and have to be accommodated.

- if the device has a distinctive set of DHCP options, we map those options to an OS name and include it in the signature. We do not use the DHCP signatures of Android or Windows as these are common across so many manufacturers as to not be useful for disambiguation. We use this mainly for embedded systems and Internet-o'-Things devices which tend to have more diverse software implementations.

- if nothing else, we will qualify the signature with the OUI of the manufacturer. This is how we distinguish the aforementioned Motorola, Sony, and Oneplus devices, and many Android devices from Samsung, LG, etc.

    We do not use the OUI for Apple devices: Apple's production volume is so large that new OUIs are added faster than we can keep up.

## Experimental results

This system has been deployed in a production network with many thousands of APs. Using a database of 382 signatures covering 132 device models, the system successfully identified 260,961 of 441,388 wifi client devices on a particular day in May 2016. This is approximately 59% of the total devices involved in the exercise. During a second exercise in late May 2016 the system successfully identified 320,769 of 549,174 wifi client devices examined, 58% of the total.

A total of 14,333 distinct signature patterns were noted during this time. This does not mean there were 14,333 different species of devices present in the overall population, as most devices have multiple signatures they have been noted to send. Some examples:
- devices capable of both 2.4GHz and 5GHz operation essentially always have a different signature on each band.
- some devices, Apple devices in particular, reduce the number of streams they offer in their HT and VHT IEs depending on noise level.

---

[1] With gratitude to Jouni Malinen for explaining why clients do this.

- most Android devices have been noted to omit their Extended Capabilities from time to time. We end up with multiple signatures in the database where some have extcap and some do not.
- OS updates sometimes change the signature, when the WPA supplicant is updated. Generally both the old software version and the new will be present within the population, resulting in multiple distinct signatures which all map to the same device.

Most of the long tail of devices are unidentified. We don't know how many distinct device species are present in the overall population, we would have to identify all of the unknown signatures in order to know how many mapped to the same device.

Of the top 50 species appearing in the sample, the signature database identified 45 of them. Of the top 100 species, 77 were identified.

## Signature History and Refinement

The first version of the signature format extracted the Information Element type ID numbers, the capabilities bitmask from the HT and VHT IEs (if present), and the Wifi Protected Setup name if present. For example, the v1 signature for several devices is shown below. Many of Google's Nexus devices include a WPS header in every Probe, which initially seemed very promising but proved not to be a widespread practice among device manufacturers.

> **Samsung Galaxy S5**
> `wifi|probe:0,1,45,127,107,191,221(506f9a,16),221(001018,2),221(00904c,51),221(00904c,4),221(0050f2,8),htcap:006f,vhtcap:0f805832|assoc:0,1,33,36,48,45,127,107,191,221(001018,2),221(00904c,4),221(0050f2,2),htcap:006f,vhtcap:0f805832`
>
> **Nexus 7 (2013 edition)**
> `wifi|probe:0,1,45,221(0050f2,8),221(0050f2,4),221(506f9a,9),htcap:016e,wps:Nexus_7|assoc:0,1,48,45,221(0050f2,2),127,htcap:016e`
>
> **iPhone 5s**
> `wifi|probe:0,1,45,127,107,221(001018,2),221(00904c,51),221(0050f2,8),htcap:0062|assoc:0,1,33,36,48,45,70,221(001018,2),221(00904c,51),221(0050f2,2),htcap:0062`

The second version of the signature format extracted more fields to disambiguate devices. It extracted the aggregation and MCS rate information from the HT and VHT IEs, and the extended capabilities bitmask. The v2 signatures for the aforementioned devices are:

> **Samsung Galaxy S5**
> `wifi2|probe:0,1,45,127,107,191,221(506f9a,16),221(001018,2),221(00904c,51),221(00904c,4),221(0050f2,8),htcap:006f,htagg:17,htmcs:000000ff,vhtcap:0f805832,vhtrxmcs:0000fffe,vhttxmcs:0000fffe,intwrk:0f,extcap:80080000|assoc:0,1,33,36,48,45,127,107,191,221(001018,2),221(00904c,4),221(0050f2,2),htcap:00`

```
6f,htagg:17,htmcs:000000ff,vhtcap:0f805832,vhtrxmcs:0000fffe,vhttxmcs:0000f
ffe,intwrk:0f,extcap:80000000
```

**Nexus 7 (2013 edition)**
```
wifi2|probe:0,1,45,221(0050f2,8),221(0050f2,4),221(506f9a,10),221(506f9a,9)
,htcap:016e,htagg:03,htmcs:000000ff,wps:Nexus_7|assoc:0,1,33,36,48,45,221(0
050f2,2),127,htcap:016e,htagg:03,htmcs:000000ff,extcap:020a0000
```

**iPhone 5s**
```
wifi2|probe:0,1,45,127,107,221(001018,2),221(00904c,51),221(0050f2,8),htcap
:0062,htagg:1a,htmcs:000000ff,intwrk:0f,extcap:04080000|assoc:0,1,33,36,48,
45,70,221(001018,2),221(00904c,51),221(0050f2,2),htcap:0062,htagg:1a,htmcs:
000000ff
```

This helped, but we still faced cases of identical signatures for similar devices. Apple devices are particularly difficult, as Apple makes a practice of using very similar hardware designs in each generation of iPhone + iPad + iPad Mini + iPod + Apple TV. Though we applaud Apple's engineering efficiency, it does make our job harder. With the v2 format there were identical signatures in a number of cases between different models within generations of Apple devices.

With far fewer ambiguous cases remaining to solve for, the v3 signature format concentrated on the devices from manufacturers which leverage similar designs in different products. Apple and Samsung are the two most prominent, and fortunately are also quite disciplined in their Wifi implementations. Both manufacturers routinely include a Power Capability IE and calibrate it for the board design. The v3 signature added both the Power Capability IE and the fixed Capabilities bitmask from the Associate Request.

The v3 signatures of our example devices are:

**Samsung Galaxy S5**
```
wifi3|probe:0,1,45,127,107,191,221(506f9a,16),221(001018,2),221(00904c,51),
221(00904c,4),221(0050f2,8),htcap:006f,htagg:17,htmcs:000000ff,vhtcap:0f805
832,vhtrxmcs:0000fffe,vhttxmcs:0000fffe,intwrk:0f,extcap:80080000|assoc:0,1
,33,36,48,45,127,107,191,221(001018,2),221(00904c,4),221(0050f2,2),htcap:00
6f,htagg:17,htmcs:000000ff,vhtcap:0f805832,vhtrxmcs:0000fffe,vhttxmcs:0000f
ffe,intwrk:0f,txpow:e20b,extcap:80080000
```

**Nexus 7 (2013 edition)**
```
wifi3|probe:0,1,45,221(0050f2,8),221(0050f2,4),221(506f9a,10),221(506f9a,9)
,htcap:016e,htagg:03,htmcs:000000ff,extcap:020a0000,wps:Nexus_7|assoc:0,1,3
3,36,48,45,221(0050f2,2),127,htcap:016e,htagg:03,htmcs:000000ff,txpow:1e0d,
extcap:020a0000
```

**iPhone 5s**
```
wifi3|probe:0,1,45,127,107,221(001018,2),221(00904c,51),221(0050f2,8),htcap
:0062,htagg:1a,htmcs:000000ff,intwrk:0f,extcap:04080000|assoc:0,1,33,36,48,
45,70,221(001018,2),221(00904c,51),221(0050f2,2),cap:0011,htcap:0062,htagg:
1a,htmcs:000000ff,txpow:1603
```

Though the v3 signature format provided more ways to disambiguate devices, it had a significant flaw: addition of the Capabilities bitmask was a mistake. There are a number of bits in the Capabilities field which depend very strongly on the environment the device is operating in and the capabilities seen in the Beacon from the AP. Addition of the "cap:" field resulted in a combinatorial explosion of signatures due to the number of variations seen from many common devices.

The v4 signature format removed the Capabilities bitmask. Additionally:
- The v4 signature format removed the Interworking field. Interworking was added in v2 because it appeared to be a good way to disambiguate Apple devices, important given Apple's practice of using very similar hardware designs in each generation of iPhone/iPad/iPad Mini/iPod/Apple TV.

    In retrospect, these differences were simply bugs which Apple later fixed. Earlier releases of Apple's software put 0xff in the Interworking IE (which is clearly not correct for a client device), later versions settled on 0x0f, while iPhone 5c populated 0x03 for reasons known only to Apple. We happened to include this field in the v2 signature during the transition.

- The v4 signature corrected the extraction of Extended Capabilities. The v3 signature implementation would only extract the first 4 bytes, and byte-swapped the field compared to the rest of the fields in the signature.

The v4 signatures of our example devices are:

**Samsung Galaxy S5**
`wifi4|probe:0,1,45,127,107,191,221(506f9a,16),221(00904c,4),221(0050f2,8),221(001018,2),htcap:006f,htagg:17,htmcs:0000ffff,vhtcap:0f815832,vhtrxmcs:0000fffa,vhttxmcs:0000fffa,extcap:0000088001400040|assoc:0,1,33,36,48,45,127,107,191,221(00904c,4),221(001018,2),221(0050f2,2),htcap:006f,htagg:17,htmcs:0000ffff,vhtcap:0f815832,vhtrxmcs:0000fffa,vhttxmcs:0000fffa,txpow:e20b,extcap:0000088001400040`

**Nexus 7 (2013 edition)**
`wifi4|probe:0,1,45,221(0050f2,8),221(0050f2,4),221(506f9a,10),221(506f9a,9),htcap:016e,htagg:03,htmcs:000000ff,extcap:00000a02,wps:Nexus_7|assoc:0,1,33,36,48,45,221(0050f2,2),127,htcap:016e,htagg:03,htmcs:000000ff,txpow:1e0d,extcap:00000a02`

**iPhone 5s**
`wifi4|probe:0,1,45,127,107,221(001018,2),221(00904c,51),221(0050f2,8),htcap:0062,htagg:1a,htmcs:000000ff,extcap:00000804|assoc:0,1,33,36,48,45,70,221(001018,2),221(00904c,51),221(0050f2,2),htcap:0062,htagg:1a,htmcs:000000ff,txpow:1603`

## Related Work in Prior Art

Jonathan Ellch [1] (also writing as Johnny Cache in [2]) proposes observation of the Duration field from the Wifi header, and shows that different chipsets and vendors have statistically significant differences in the choice of values they populate.

Franklin, McCoy, Tabriz, Neagoe, Van Randwyk, and Sicker [3] observe cyclical patterns in the timing of Probe requests sent by the client, and show that different chipsets and vendors have distinguishable patterns in their timing.

Bratus, Cornelius, Kotz, and Peebles [4] propose active techniques, testing a driver's response to various malformed or underspecified fields in management frames.

Gopinath, Bhagwat, and Gopinath [5] show that MAC-level implementation details such as back-off timers vary substantially between vendors, and that this can be used to distinguish chipsets via passive observation.

Pang, Greenstein, Gummadi, Seshan, and Wetherall [6] show that the content and timing of traffic sent by Wifi clients can often distinguish individual users even if the MAC address of their device is replaced with a pseudonym address.

Previously documented techniques in this area are mostly from the pre-802.11n era. The Wifi standards of that time, such as 802.11g, made relatively little use of optional fields and parameters in Wifi MAC Layer Management Entity (MLME) frames. The techniques of that time therefore focussed on other mechanisms like timing thresholds chosen by the driver. The MLME-based approach proposed in this paper would not work well in a population of 802.11a/b/g devices with relatively little variation in their management frame contents.

More recently Neumann, Heen, and Onno [7] showed that passive analysis of factors such as packet inter-arrival times can distinguish individual users, and that these factors would be difficult for an adversary to obfuscate. The MLME mechanism proposed here would be relatively easy for an attacker with control of the WPA supplicant implementation to defeat.

Additionally, the pioneering work on DHCP signatures implemented by the Fingerbank project (http://www.fingerbank.org/) [8] provided the inspiration for the signature method described here and also serves as a supplemental source of OS identification information for cases where the Wifi signature is indistinct.

## Future Work: AP Taxonomy

A very similar mechanism could be applied to Beacon frames, to identify the models of nearby APs. As with the Probe and Association Request frames described in this paper, Beacons have become substantially more diverse in recent years with the addition of optional fields and vendor extensions. This is especially true among APs designed for the Enterprise market, but even consumer grade APs often include vendor-specific IEs intended to coordinate with that same vendor's client chipsets and drivers.

## Future Work: Storing Probe signatures

The signature described in this proposal uses the contents of both the Probe and the Associate Request from the client. Our implementation in hostapd stores the signature information in the station info structure for associated clients, which means that the information from Probes sent by that client prior to its association are not retained. We need to see the client associate, when the Associate Request will be stored, and then we need to see *another* Probe Request from the client before the signature can be output.

This is generally not an issue, as client devices send lots of Probes. Lots and lots of Probes, actually. However there are a few devices which are very sparing in the Probes they send. Notably, ChromeOS devices tend to send a Probe to find nearby APs, an Associate Request to join one of those APs, and then no more Probes for an extended period of time. The current implementation can take quite a long time before it outputs a signature for devices which exhibit this behavior.

A future enhancement would be to store Probe information for clients which have not yet associated with the AP. This data structure would need to be pruned aggressively, as there can be very large numbers of Probes from devices which are just passing through and will never attempt to Associate.

## Epilogue
or, *What We Learned By Staring At Wireshark For Far Too Long*

Wifi client devices are very silly creatures. Some behaviors we have noticed while working on this mechanism:
- Qualcomm-Atheros chipsets have a proprietary feature where they can use 802.11ac MCS rates in the 2.4 GHz band. When sending a Probe Request for the Broadcast SSID, devices with a QCA chipset often include a VHT Capabilities IE even though 802.11ac defines VHT only for 5GHz operation.

    When sending a directed Probe to the SSID of our AP these devices do not include a VHT Capabilities IE. We assume that they look for VHT information in an AP's Beacon, and omit their own VHT Capabilities from the directed Probe if not seen.

    Intel 7260 chipsets were also noted including VHT Capabilities in their 2.4GHz Probe, though the Intel chipsets include the VHT information in all Probes both Broadcast and specific.

- Broadcom also claims to have proprietary support for 802.11ac MCS rates in the 2.4GHz band, but is much more stealthy about it. We did not notice unusual content in 2.4GHz Probes sent by Broadcom chipsets, but we do note that Broadcom's vendor extension 221(001018,2) is one of the most common vendor IEs we've seen. This IE contains what looks like a 6 byte bitmask, which would cover a lot of capabilities. There may be a magic *VHT-for-2.4GHz* bit in there somewhere.

- When you add Spectrum Management support to your AP and start setting the RRM bits in your Beacons, you get to watch in horror as about half of the signatures you've collected suddenly change when IE #70 (Radio Resource Management) appears.

- Many devices include a Current Channel IE (type #3) when sending a directed Probe for a specific SSID, but not when sending a Broadcast Probe to find all nearby APs. This is a bit inexplicable: Broadcast Probes are often sent as part of a scan through all available channels, where including the Current Channel IE seems somewhat more useful than when sending to just one channel.

    We assume that stack developers want to format the Broadcast Probe in memory just once and not have to change the IE contents each time they move on to the next channel to scan.

- During standardization of 802.11n vendors wanted to ship products to market sooner, and adopted a "Pre-N" marketing term for interoperable but pre-standard products. In Wifi MLME frames this took the form of a vendor IE from Epigram, Inc, which strongly resembles the later standardized 802.11n HT Capabilities IE.

    In an iOS version sometime around the release of the iPhone 6, Apple stopped including the Pre-N vendor IE from Epigram, Inc and only includes the standard HT Capabilities IE. Thus, there is an existence proof that it is possible to eventually phase out marketing efforts like this. It takes about 10 years.

The database of signatures developed as part of this effort can be found at:
https://github.com/NetworkDeviceTaxonomy/wifi_taxonomy